\documentclass[usenatbib,conference]{basi}
\usepackage[T1]{fontenc}
\usepackage[british]{babel}
\usepackage[varg]{txfonts}
\usepackage{rotating}
\usepackage{dcolumn}
\begin{document}
\title[M-dwarfs and colder with VOSA]{Physical parameters of young M-type stars and brown dwarfs with VOSA\thanks{
	%By Amelia Bayo, email: \texttt{bayo@mpia.de}}}
            %, or see
            %
            \texttt{http://svo2.cab.inta-csic.es/svo/theory/vosa/}}}
\author[A.~Bayo et~al.]%
       {A.~Bayo$^{1,2}$,
       C.~Rodrigo$^{3,4}$,
%       D.~Barrado$^{3}$
       D.~Barrado$^{3}$
       E.~Solano$^{3,4}$,
       F.~Allard$^5$,
        and V.~Joergens$^{1}$\\
       $^1$Max Planck Institut f\"ur Astronomie, K\"onigstuhl 17, 69117, Heidelberg, Germany\\
       $^2$European Southern Observatory, Alonso de C\'ordova 3107, Vitacura, Santiago, Chile\\
       $^3$Depto. Astrof\'isica, Centro de Astrobiolog\'ia (INTA-CSIC), P. O. Box 78, E-28691 Villanueva de la Ca\~nada, Spain\\
       $^4$Spanish Virtual Observatory, Spain\\
       $^5$Centre de Recherche Astronomique de Lyon (CRAL), \'Ecole Normale Sup\'erieure de Lyon, 69364, Lyon, France}

\pubyear{2014}
\volume{9}
\pagerange{\pageref{firstpage}--\pageref{lastpage}}
%\status{submitted}

\date{Received --- ; accepted ---}

\maketitle
%------------------------------------------------------------------------------%
% abstract and keywords                                                        %
%------------------------------------------------------------------------------%
\label{firstpage}

\begin{abstract}
{\small Although M dwarfs are the most common stars in our stellar neighborhood they are still among the least understood. This class of objects is dominated by dramatic changes: in their interiors (fully convective, with implications in angular momentum evolution), in their atmospheres (crossing temperatures where dust settling occurs), and in their nature (including both, stellar and substellar objects).\vspace{-0.1cm}

Populating efficiently our solar neighborhood,  they are very well represented in the databases coming from new and more sensitive surveys that provide photometry at many wavelength ranges and cover large areas of the sky (few examples among many others are GALEX, SDSS, 2MASS, WISE and VISTA). \vspace{-0.1cm}

In this context of opulence of objects and data, the Virtual Observatory comes in naturally as an excellent framework to efficiently advance in the knowledge of M-type sources. We put special emphasis in the benefits of using the new capabilities of VOSA (Virtual Observatory SED Analyzer, \citealt{2008A&A...492..277B}; in operation since 2008 and in constant development) to study large samples of candidate and confirmed M members of Chamaeleon I.\vspace{-0.1cm}

%This illustrates the usage of various features of Version 2.1 of the \LaTeX\
%class file for papers for the Bulletin of the Astronomical Society of India.
%The class file \verb|basi.cls|, this sample paper and another guide are
%available from the BASI webpage:\\[6pt]
%
%\hbox to 30pt{\hfil}\verb|VOSA: http://svo2.cab.inta-csic.es/svo/theory/vosa/|
}
\end{abstract}


%------------------------------------------------------------------------------%
% main text of the paper, using \section, \subsection, \subsubsection          %
%------------------------------------------------------------------------------%
\section{Introduction}\label{s:intro}

\vspace{-0.1cm}The determination of stellar (and substellar) parameters is by no means a new task in astronomy and many different approaches, methods, etc. can be found in the literature. However, the data avalanche that we are experiencing and that will continue in the near and far future should make us reconsider the efficiency of classical methodologies and how to improve them.\vspace{-0.2cm}

\begin{figure}
\includegraphics[width=12cm]{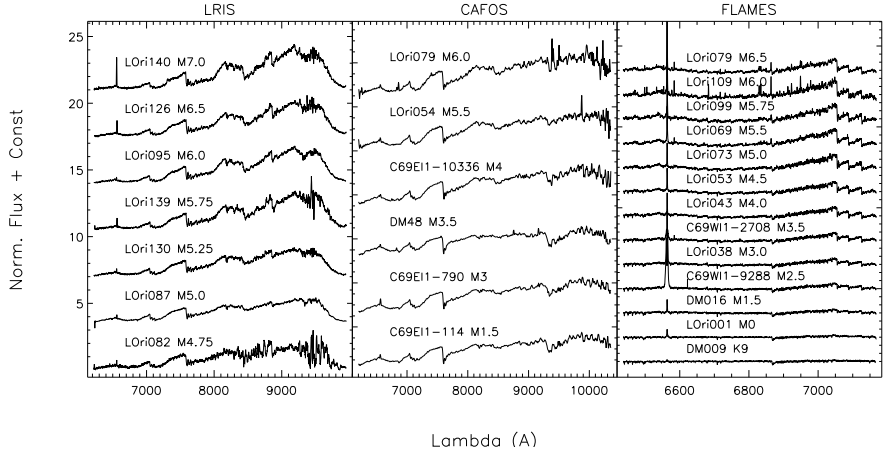}
\caption{Spectral templates for young late-type stars (members of the $\sim$5-12 Myr cluster Collinder69) presented in \citet{2011A&A...536A..63B} covering a wide range of spectral resolutions.}
\label{specseq}
\vspace{-0.2cm}
\end{figure}

A zero order classification of these approaches can originate on whether the method is based on empirical data or theoretical models. There is a number of advantages and disadvantages for both cases: for example determining effective temperatures and gravities using templates (Fig.~\ref{specseq}) or spectral indices is model independent, but relies strongly not only in the availability of benchmark stars with similar parameters than the ones under study, but also on the particular set of indexes and/or templates being suitable for the resolution and wavelength coverage of the observations of the targets to be analyzed. In addition, these comparisons allow us to build classes of objects but does not necessarily provide great insight in the physics of the objects we are classifying (especially when the actual physics of the objects, like M-type stars, is not well understood yet).\vspace{-0.2cm}

On the other hand, the comparison with models does not have the intrinsic problems of observations in terms of resolution, wavelength coverage, etc. because one can start with a model of resolution as high as wanted and degrade, convolve, etc. the model to simulate how it would be seen with the same observational set-up as the data to be analyzed. However, obviously, this approach is, by nature, completely model dependent and comparisons with different models are desirable until a deep understanding of the physics of the objects under study is achieved.\vspace{-0.2cm}

Any of the two approaches poses a clear challenge nowadays when multi-wavelength observations are publicly available for an immense number of objects: how do we efficiently combine our own private data with those taken by different consortia, with a variety of instruments, techniques,...? In addition to this, when following the model comparison approach, one has to add the extra-step of incorporating different kinds of models and customize them to the same peculiarities than the observations.\vspace{-0.2cm}

The Virtual Observatory (VO) provides good means to overcome most of these challenges: during almost a decade, the VO community has pursued the goal of defining standards and software that ease the task of combining private, archival, and theoretical data, making this process transparent to the astronomer.\vspace{-0.2cm}

As part of these efforts, the Spanish Virtual Observatory developed VOSA (Virtual Observatory SED Analyzer, \citealt{2008A&A...492..277B}), and we have kept on improving its capabilities ever since. These improvements include the large effort of creating a VO-compliant filter profile service (an homogeneous database with exhaustive description of the filters available in most of the main ground and space-based observatories), key-point in the structure of VOSA and extremely useful for many different kinds of projects besides VOSA. In summary, VOSA combines private photometric measurements with data available in the VO, builds spectral energy distributions (SEDs), accesses various theoretical models to simulate the equivalent theoretical SEDs, allows the user to decide the range of parameters to explore, performs the SED comparisons (not only with the synthetic SEDs but also with synthetic photometry from templates), provides the best fitting models to the user following two different (complementary) approaches, and, finally, compares these parameters with isochrones and evolutionary tracks to estimate masses and ages.\vspace{-0.2cm}

In the following we briefly describe three sets of objects that we have studied. In Section 3, we also provide more details on the approach followed by VOSA (although we refer the reader to the very throughout description in Bayo et al. 2013a, submitted). In Section 4, we discuss the results obtained and in Section 5, we briefly comment on the future plans for the tool.\vspace{-0.2cm}

%\vspace{-0.1cm}
\section{The samples}

\vspace{-0.1cm}To illustrate the contribution of VOSA to the study of young M-type sources we have selected two sets of objects in the Chamaeleon I (Cha I) region of the sky: on the one hand a sample of 11 confirmed low-mass stars (VLMs) and brown dwarfs (BDs) members of the dark cloud \citep{2000A&A...359..269C} and, on the other hand, a large sample of Class III candidate members to Cha I.\vspace{-0.2cm}

The first sample of objects covers the M6 -- M8 spectral range (with IDs Cha H$\alpha$1 - 9, Cha H$\alpha$ 11 and Cha H$\alpha$ 12), a mix of (young, $\sim$2 Myr) very low mass stars and brown dwarfs with the adequate temperatures for dust settling to happen. For these objects we have our own optical and near-infrared low resolution spectroscopy that will be used as sanity checks to the SED fits quality. We added to the sample three members of the TW-Hydra association (TWA) extending the spectral type sequence down to M8.5 (2MASS J1139511-315921, SSSPMJ1102- 3431 and 2MASS J1207334-393254) to have a better coverage of the temperature range for dust settling.\vspace{-0.2cm}
% where the dust settling in the For the first sample of objects we have optical and near-infrared spectroscopy that will serve as quality check for the SED-fitting derived parameters. In addition, to complete the sequence of late M-type young sources for which we had our own spectroscopic data, we have added three brown dwarfs, confirmed members of the TWA moving group.%to the first sample to which we add three brown dwarfs from the TWA moving group to have a complete seque

The second sample of objects (diskless candidate members to Cha I) was selected imposing very basic and simple criteria: we queried VizieR \citep{2000A&AS..143...23O} for the WISE objects within 2$^{\circ}$ (about 5.8 pc at the cloud distance) of the rough center of the cloud
% ($11^{\rm h}07^{\rm m}00^{\rm s}  -77^{\rm o}59'00"$)
, with good quality flag photometry and fulfilling $|W1-W2|\le$ 0.25 mag \& $|W3-W4|\le$ 1.0 mag. The latter cuts define the space compatible with main sequence and weak-line TTauri stars (see as an example, \citealt{2012ApJ...751..114S}). This query returned a VOTable (standard format for tabular data in the VO) containing 1634 sources.% with their respective WISE photometry.% that conformed the second sample for VOSA analysis.% (the WISE Class III initial candidate sample); the spatial distribution of the objects is shown in Fig. 10 with the WISE W1 mosaic as background.
\vspace{-0.2cm}

Finally, to demonstrate how VOSA can also be very efficient in studying individual (in this case exotic) objects, %although it was conceived to work with large samples of objects,
 we show the role it played in the characterization of the lowest accreting object known to date: OTS 44 (also a member of the Chamaeleon I dark cloud with an estimated mass of $\sim$12 M$_{\rm Jup}$, \citealt{2013A&A...558L...7J}), for which, besides publicly available photometry, we had a low resolution optical spectrum.\vspace{-0.2cm}

%Three different samples: the dust settling one, the possible class II sources and the extremely low mass BD or IMPO

\section{The Method: VOSA, new capabilities}

\vspace{-0.23cm}This section aims to be a description on the capabilities of VOSA for stellar cases and how to use the tool. For consistency checks and further discussion in technical aspects of the tool, we refer the reader to Bayo et al. (2013a, submitted) and \citet{2008A&A...492..277B}, where the first version of VOSA was presented. On the other hand, the galaxies case is developed in \cite{2011IAUS..277..230S}. \vspace{-0.2cm}

In Fig. 2, we display a snapshot of the upper-part of the VOSA web-interface. The basic idea behind the tool is that it provides different tasks that the user can perform, and the natural way to complete them is from left to right (although most tasks are optional and it is up to the user to complete them or not).\vspace{-0.cm}

\begin{figure}
\includegraphics[width=12cm]{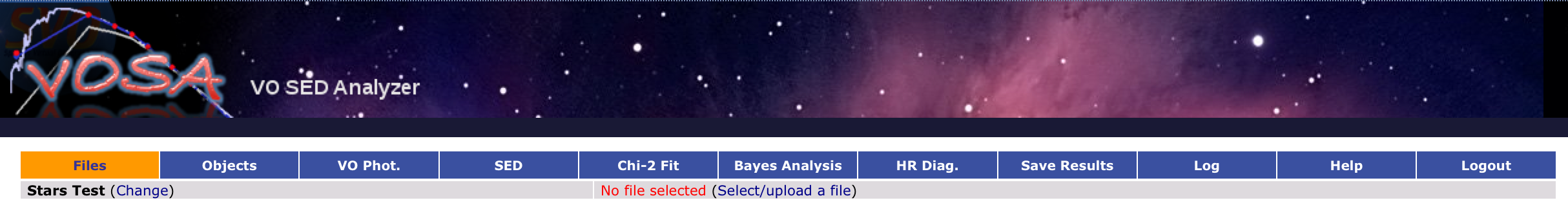}
\caption{Tab-like structure of VOSA. The tool is designed so that to complete a workflow, the user follows the tabs corresponding to the tasks to be completed from left to right (the tab that is in usage is highlighted in orange).}
\label{vosa}
\end{figure}\vspace{-0.2cm}

We will briefly walk the reader through a complete workflow: \vspace{-0.2cm}

\noindent - In the ``Files" tab the project is defined. Private data such as photometry, distance, extinction, etc. can be ingested (using a format converter from csv or VOTable input), and also notes can be kept on the details of the project. We note that these projects can consist of an individual object (typing the coordinates of the object or a Sesame\footnote{http://cds.u-strasbg.fr/cgi-bin/Sesame} resolvable ID) or a file with hundreds or even thousands of objects. %Regarding the characteristics of the photometry uploaded, the requirement is that the photometric system and instrument
\vspace{-0.25cm}

\noindent - ``Objects" tab: non-photometric information (coordinates, distances and extinction values) is searched in the VO within the requested radius for each object under study.\vspace{-0.25cm}

\noindent - ``VO Phot." tab: a large selection of catalogs (including: 2MASS, IRAS, GLIMPSE, C2D, WISE, VISTA surveys, Tycho-2, etc.) is offered covering wavelengths from the ultraviolet to the far infrared (IR). The query for counterparts for each catalog can be customized in terms of magnitude and radius filters. \vspace{-0.25cm}

\noindent - ``SED" tab: the built SEDs (including private and archival data) can be visualized and ``manipulated" in different ways: on the one hand, all data points are displayed including quality flags (when available in the parental catalog) and ``point properties", where the user can choose to include one particular point or not in the fit or to highlight it as an upper-limit. On the other hand, VOSA automatically looks for IR excess and marks those points as ``not-to be fitted" when photospheric models are used. If the detected excess is not adequate for the user's purpose, this excess filter can also be modified (object-by-object or all sources at once). \vspace{-0.25cm}

\noindent - ``Chi-2 Fit" tab: a selection of models is offered (from hot-intermediate and cold photospheres to C and O-rich evolved objects) for the fit. Once the models have been selected, the parameter space to be probed is requested and VOSA builds the synthetic SEDs within the parameter space for direct comparison with the observations. VOSA returns the best five fits per family of models in terms of minimum $\chi^2$ and uses the best fitting model to perform a panchromatic interpolation (infer the flux of the objects for the wavelength ranges where no observational data are available). The best-fit results table includes T$_{\rm eff}$, A$_{V}$, L$_{\rm bol}$, $\lambda$ at which the IR excess was detected, $\log(g)$, etc. and can be sent with one click, to other VO-tools such as TOPCAT, VOStat, etc.\vspace{-0.25cm}

\noindent - ``Bayes Analysis" tab: the probability distribution function (PDF) of each fitted parameter is calculated so that not only the minimum $\chi^2$ fit is provided but also the significance of the explored parameters. \vspace{-0.25cm}

\noindent - ``HR Diag." tab: the estimated T$_{\rm eff}$ and L$_{\rm bol}$ pairs are displayed in a Hertzsprung-Russell diagram where several collections of isochrones and evolutionary tracks can also be displayed and masses and ages are estimated via their interpolation. The results from the HR diagram can also be sent to other VO-tools.\vspace{-0.25cm}

\noindent - ``Save Result" tab: used to save all the VOSA products (complete SEDs, best fitting models, PDFs, etc.) with several tabular and graphics formats. In addition to the VOSA products, a ``.bib" file with the adequate references to all the resources used (catalogues queried, theoretical models, etc.) is provided to ease proper referencing.\vspace{-0.25cm}

\noindent - The ``Log" tab can be consulted at any time within the workflow to recover the parameters used in searches, models, etc. \vspace{-0.25cm}

\noindent - The ``Help" tab provides detailed explanations about all the tasks, formats, algorithms, etc. used by VOSA.\vspace{-0.25cm}

\section{Results}

\subsection{Impact of dust treatment in the models for parameter estimations}

\vspace{-0.1cm}We followed the above described workflow for the sample of late-M VLMS and BDs (belonging to either Cha I or TWA) using model collections based on three different approaches for the dust properties: BT-Settl (\citealt{2012RSPTA.370.2765A}, integral treatment), COND (\citealt{2003IAUS..211..325A}, total gravitational settling) and DUSTY (\citealt{2003IAUS..211..325A}, inefficient gravitational settling)\vspace{-0.2cm}

In Fig. 3, we show the comparison of the results from the three models. Clearly, the differences in the estimated parameters increase with decreasing effective temperature yielding unrealistic values for the two limiting cases (COND and DUSTY) for temperatures lower than 2500 K. On the other hand, the BT-Settl models provide parameters in good agreement with the literature for the latest spectral types (the TWA members) and some discrepancies in T$_{\rm eff}$ of the Cha I sample discussed below.\vspace{-0.5cm}% previous approached \citep{Allard03, Allard12}

%\vspace{-0.3cm}
\begin{figure}
\centerline{\includegraphics[width=6.5cm]{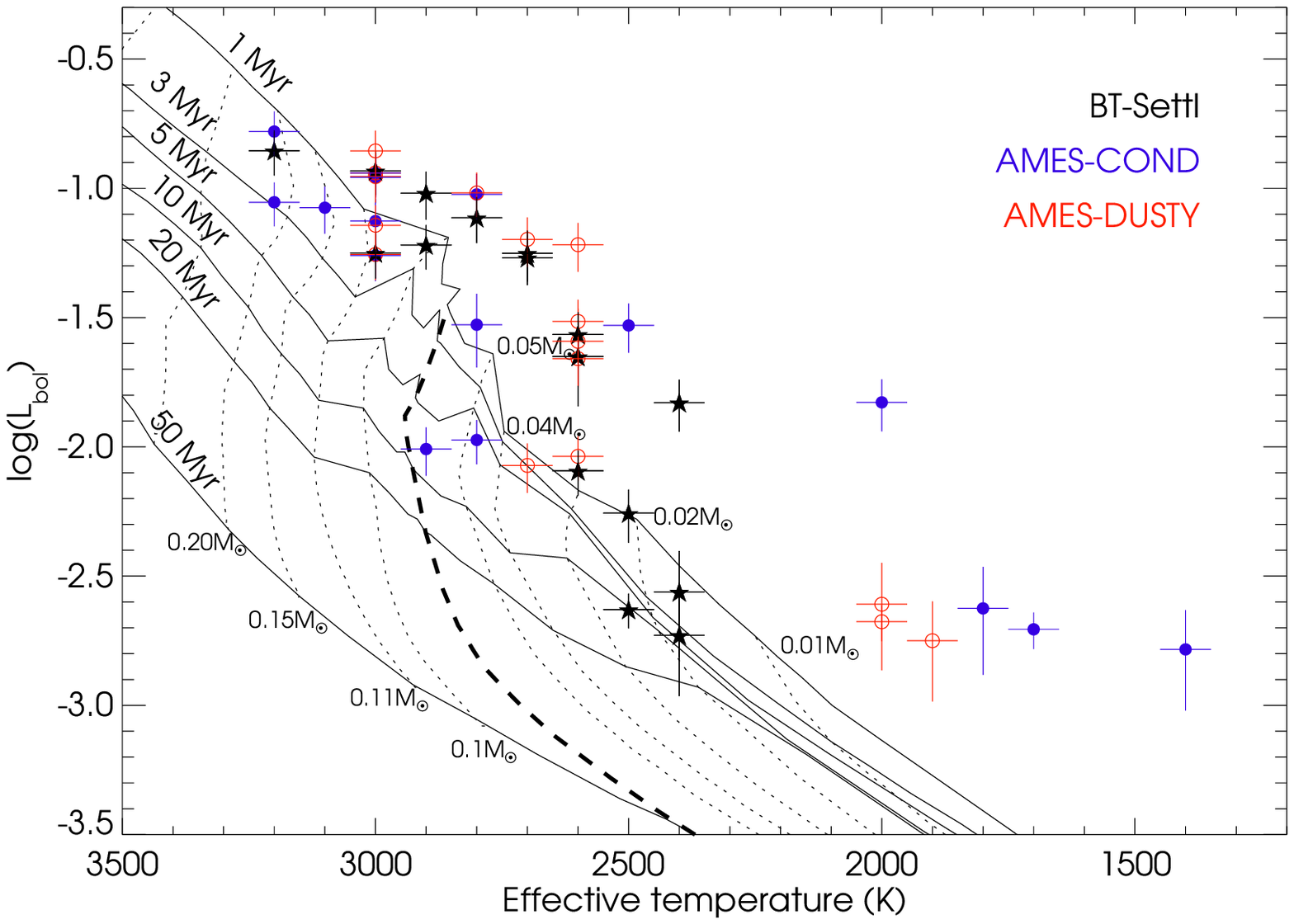} 
            \includegraphics[width=6.5cm]{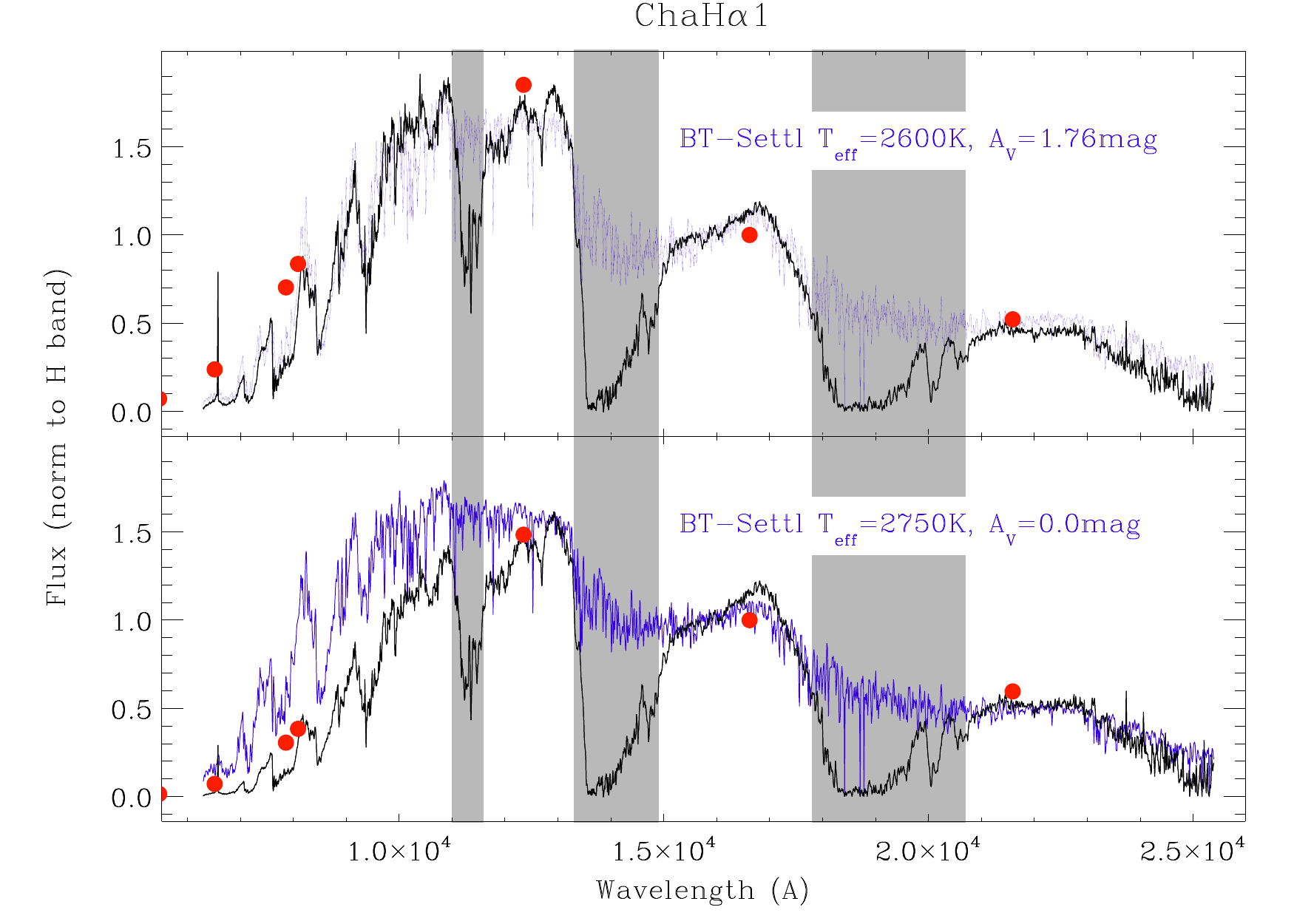}}
%\medskip
%\centerline{\includegraphics[angle=90,width=4cm]{asi-logo.eps} \qquad
%            \includegraphics[origin=lt,angle=270,width=4cm]{asi-logo.eps}}
%
\caption{\textbf{Left:} HR-diagram showing the results achieved with the different dust treatments. \textbf{Right:} Comparison of the parameters obtained with VOSA and the BT-Settl models (grey areas to be ignored due to lack of atmospheric transmission, black line for the observe spectra and blue for the model; red dots for the photometry) with those reported in \cite{2007ApJS..173..104L}. Figures from Bayo et al 2013b, submitted.\label{dust_teffScale}} 
\end{figure}

\begin{figure}
\centerline{\includegraphics[width=6.7cm]{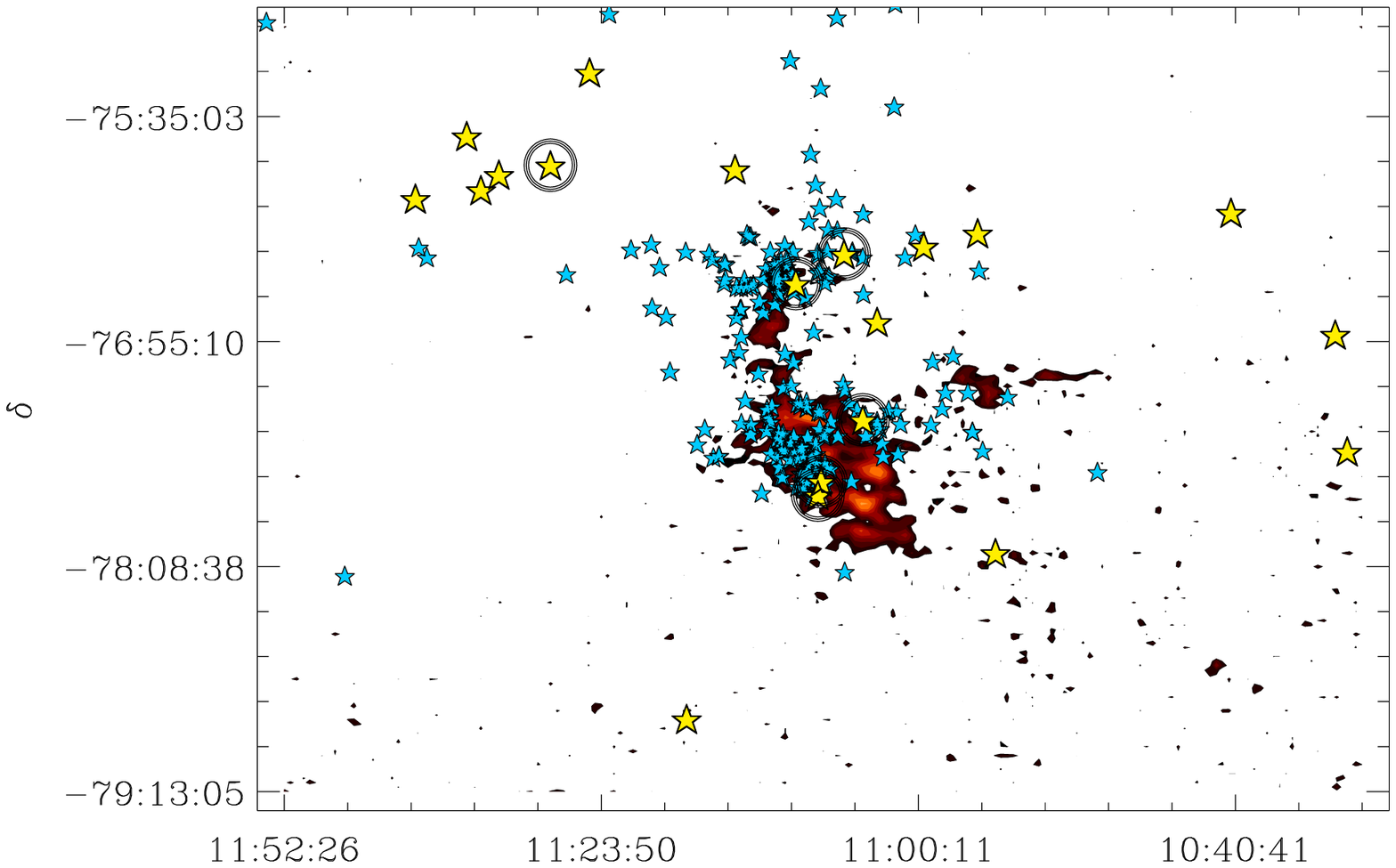} 
            \includegraphics[width=5.5cm]{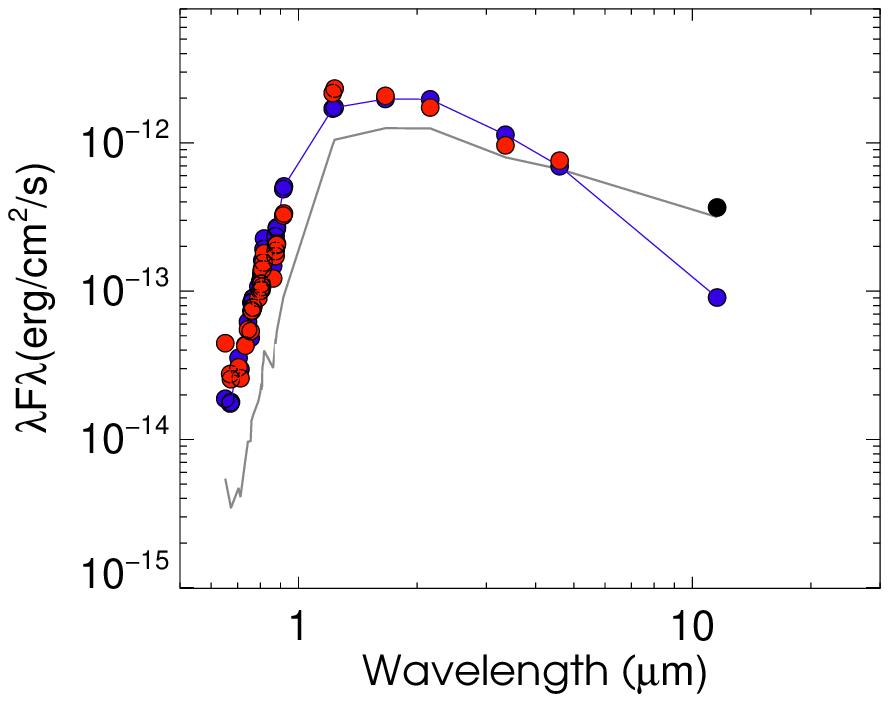}}
%\medskip
%\centerline{\includegraphics[angle=90,width=4cm]{asi-logo.eps} \qquad
%            \includegraphics[origin=lt,angle=270,width=4cm]{asi-logo.eps}}
%
\caption{\textbf{Left:} Spatial location of the new Class III candidates (yellow stars) with respect to the known members (light-blue stars) on top of a color scale extinction map. \textbf{Right:} Photospheric VOSA-fit for the extremely low-mass BD / planetary mass object OT44. Gray line for the ``raw" enriched SED, red and black dots for the de-reddened one (black highlights infrared excess) and blue line-connected dots for the best fitting model.\label{dust_teffScale}} 
\end{figure}

\subsection{Differences with previously reported temperature scales}

\vspace{-0.1cm}As mentioned before, the best-fitting BT-Settl models for the Cha I sample suggest lower temperatures than those reported in \cite{2007ApJS..173..104L}. This is not surprising since the latter work used the obsolete (for low temperatures) Next-Gen grid where no dust settlement treatment is included. As sanity check, in Fig. 3, we show how, once we include the optical and NIR spectra, the theoretical model corresponding to the VOSA-fit reproduces the continuum and features better than that resulting from assuming the parameters given in \cite{2007ApJS..173..104L}.\vspace{-0.5cm}

\subsection{Discovering new candidate members to Chamaleon I}

\vspace{-0.3cm}We applied the whole VOSA workflow to the large sample of excess-less sources described in Section 2. After filtering out objects with bad-quality fits and/or a location in the HR diagram that does not agree with membership to Cha I, we recovered a large fraction of the known weak-line TTauri population of Cha I and a sample of  20 objects not previously associated with the dark cloud. We are in the process of spectroscopically confirming their membership to Cha I. This confirmation would be very interesting when studying the dynamical evolution of Cha I since, as can be seen in Fig. 4, these Class III candidates are more spread in space than the known members.\vspace{-0.5cm}

\subsection{Progress in the characterization of extremely low mass objects}

\vspace{-0.1cm}VOSA and the SVO-filter profile service\footnote{http://svo2.cab.inta-csic.es/theory/fps/} were used to enrich and fit the SED from the extremely low mass accreting object OTS44 \citep{2013A&A...558L...7J}. We discretized the private low-resolution optical spectrum available and derived the BT-Settl best fit (see Fig. 4) to serve as input for the radiative transfer modeling of the disk. Given the large number of parameters involved in disk modeling, accurate knowledge of the photosphere is a critical aspect in these studies.\vspace{-0.2cm}

\section{Future plans}

\vspace{-0.1cm}As mentioned before, VOSA is in constant development while, at the same time, we try to keep it as ``general purpose" as possible. An example of the latter is how the same machinery has been adapted to the analysis of nearby galaxies (within the AVOCADO project; \citealt{2011IAUS..277..230S}).

Some of the new VOSA capabilities that we are working/planing on including:

\noindent - continue adding access to more collections of observational and theoretical data;\vspace{-0.2cm}

\noindent - finish the implementation of the extinction maps from \cite{2009A&A...508L..35K} so that the user can decide to assume those values as maximum A$_{\rm V}$ limit in the SED-fitting process;\vspace{-0.2cm}

\noindent - include proper motion treatment to guarantee consistent counterparts when cross-matching private and public data;\vspace{-0.2cm}

\noindent - provide a batch mode option for very large queries. Nowadays, depending the connection and load of the server, it takes $\sim$2 hr. to do the full workflow for $\sim$2000 objects (i.e., from ingesting your data to retrieving the results of the HR diagram). For most of this time the user interaction is not needed at all, but the process is sensitive, for example, to internet interruptions. Since the community would like to be able to analyze hundreds of thousands of objects per session, in the future we will provide a way to put the job remotely in VOSA and be notified when the analysis is finished.\vspace{-0.2cm}

\section*{Acknowledgements}
\vspace{-0.3cm}
{\small 
This research has made use of the SIMBAD database and Aladin, operated at CDS, Strasbourg, France; and the NASA's Astrophysics Data System.
This publication makes use of VOSA, developed under the Spanish Virtual Observatory project supported from the Spanish MICINN through grants AYA2008-02156 and AYA2011-24052. This work was co-funded under the Marie Curie Actions of the European Commission (FP7-COFUND), Spanish grants AYA2012-38897-C02-01, AYA2010-21161-C02-02
This publication also makes use of data products from the Wide-field Infrared Survey Explorer, which is a joint project of the University of California, Los Angeles, and the Jet Propulsion Laboratory/California Institute of Technology, funded by the National Aeronautics and Space Administration.}

%Here is where acknowledgements go, in an un-numbered section (which is
%specified by using the \verb|\section*{Acknowledgements}| command).

%------------------------------------------------------------------------------%
% bibliography: produced from ADS using custom format of                       %
%                                                                              %
%     %z132 \\bibitem[%\2%(y)%\3m]%{R}\n   %\8.1g,%\Y,%\q,%\V,%\ p             %
%------------------------------------------------------------------------------%
%\small

%\appendix
%%------------------------------------------------------------------------------%
%% appendices:                                                                  %
%%------------------------------------------------------------------------------%
%\section{An Example Appendix}\label{a:example}

%This is an example appendix, which uses the \verb|\appendix| command, and then
%subsequently the \verb|\section{...}|, \verb|\subsection{...}| etc.\ commands.

%\label{lastpage}
%------------------------------------------------------------------------------%
\end{document}